
\input harvmac
\input tables

\overfullrule=0pt
 
\def\A{{\scriptscriptstyle A}}
\def\B{{\scriptscriptstyle B}}
\def\C{{\scriptscriptstyle C}}

\def\I{{\scriptscriptstyle I}}
\def\J{{\scriptscriptstyle J}}

\def\R{{\scriptscriptstyle R}}
\def\S{{\scriptscriptstyle S}}
\def\T{{\scriptscriptstyle T}}

\def\V{{\scriptscriptstyle V}}

\def\X{{\scriptscriptstyle X}}
\def\Y{{\scriptscriptstyle Y}}


\def\CN{{\cal N}}

 
\def\a{\alpha}
\def\b{\beta}

\def\e{\epsilon}

\def\s{\sigma}


\def\half{{1 \over 2}}

\def\third{{1 \over 3}}


\def\bar#1{\overline{#1}}
\def\bzero{{b_0}}
\def\ccdot{\hbox{\kern-.1em$\cdot$\kern-.1em}}

\def\Gdual{{\widetilde G}}
\def\gtap{\raise.3ex\hbox{$>$\kern-.75em\lower1ex\hbox{$\sim$}}}

\def\Hdual{{\widetilde H}}

\def\Lambdab{{\Lambda^\bzero}}

\def\ltap{\raise.3ex\hbox{$<$\kern-.75em\lower1ex\hbox{$\sim$}}}
\def\Nc{N_c}
\def\Nf{{N_f}}
\def\R{{\scriptscriptstyle R}}
\def\qbar{{\bar{q}}}
\def\qhat{{\hat{q}}}

\def\qp{{q'}}
\def\qpp{{q''}}

\def\shat{\hat{s}}
\def\sigone{\sigma_1}
\def\sigtwo{\sigma_2}
\def\sigthree{\sigma_3}
\def\sp{\>\>}

\def\therefore{{\hbox{..}\kern-.43em \raise.5ex \hbox{.}}\>\>}

\def\Vslash{V\hskip-0.75 em / \hskip+0.30 em}
\def\wtilde{{\widetilde w}^a}
\def\Wmag{W_{\rm mag}}
\def\Wtree{W_{\rm tree}}
\def\xrm{{\rm \X}}

\newdimen\pmboffset
\pmboffset 0.022em
\def\oldpmb#1{\setbox0=\hbox{#1}%
 \copy0\kern-\wd0
 \kern\pmboffset\raise 1.732\pmboffset\copy0\kern-\wd0
 \kern\pmboffset\box0}
\def\pmb#1{\mathchoice{\oldpmb{$\displaystyle#1$}}{\oldpmb{$\textstyle#1$}}
      {\oldpmb{$\scriptstyle#1$}}{\oldpmb{$\scriptscriptstyle#1$}}}


\def\fund{  \> {\vcenter  {\vbox  
              {\hrule height.6pt
               \hbox {\vrule width.6pt  height5pt  
                      \kern5pt 
                      \vrule width.6pt  height5pt }
               \hrule height.6pt}
                         }
                   }
           \>\> }

\def\antifund{  \> \overline{ {\vcenter  {\vbox  
              {\hrule height.6pt
               \hbox {\vrule width.6pt  height5pt  
                      \kern5pt 
                      \vrule width.6pt  height5pt }
               \hrule height.6pt}
                         }
                   } }
           \>\> }

\def\sym{  \> {\vcenter  {\vbox  
              {\hrule height.6pt
               \hbox {\vrule width.6pt  height5pt  
                      \kern5pt 
                      \vrule width.6pt  height5pt 
                      \kern5pt
                      \vrule width.6pt height5pt}
               \hrule height.6pt}
                         }
              }
           \>\> }

\def\symbar{  \> \overline{ {\vcenter  {\vbox  
              {\hrule height.6pt
               \hbox {\vrule width.6pt  height5pt  
                      \kern5pt 
                      \vrule width.6pt  height5pt 
                      \kern5pt
                      \vrule width.6pt height5pt}
               \hrule height.6pt}
                         }
              }
           } \>\> }

\def\anti{ \> {\vcenter  {\vbox  
              {\hrule height.6pt
               \hbox {\vrule width.6pt  height5pt  
                      \kern5pt 
                      \vrule width.6pt  height5pt }
               \hrule height.6pt
               \hbox {\vrule width.6pt  height5pt  
                      \kern5pt 
                      \vrule width.6pt  height5pt }
               \hrule height.6pt}
                         }
              }
           \>\> }

\def\antithree{ \> 
              {\vcenter  {\vbox  
              {\hrule height.6pt
               \hbox {\vrule width.6pt  height5pt  
                      \kern5pt 
                      \vrule width.6pt  height5pt }
               \hrule height.6pt
               \hbox {\vrule width.6pt  height5pt  
                      \kern5pt 
                      \vrule width.6pt  height5pt }
               \hrule height.6pt
               \hbox {\vrule width.6pt  height5pt  
                      \kern5pt 
                      \vrule width.6pt  height5pt }
               \hrule height.6pt}
                         }
              }
           \>\> }

\def\antifour{ \> 
              {\vcenter  {\vbox  
              {\hrule height.6pt
               \hbox {\vrule width.6pt  height5pt  
                      \kern5pt 
                      \vrule width.6pt  height5pt }
               \hrule height.6pt
               \hbox {\vrule width.6pt  height5pt  
                      \kern5pt 
                      \vrule width.6pt  height5pt }
               \hrule height.6pt
               \hbox {\vrule width.6pt  height5pt  
                      \kern5pt 
                      \vrule width.6pt  height5pt }
               \hrule height.6pt
               \hbox {\vrule width.6pt  height5pt  
                      \kern5pt 
                      \vrule width.6pt  height5pt }
               \hrule height.6pt}
                         }
              }
           \>\> }

\def\antifive{ \> 
              {\vcenter  {\vbox  
              {\hrule height.6pt
               \hbox {\vrule width.6pt  height5pt  
                      \kern5pt 
                      \vrule width.6pt  height5pt }
               \hrule height.6pt
               \hbox {\vrule width.6pt  height5pt  
                      \kern5pt 
                      \vrule width.6pt  height5pt }
               \hrule height.6pt
               \hbox {\vrule width.6pt  height5pt  
                      \kern5pt 
                      \vrule width.6pt  height5pt }
               \hrule height.6pt
               \hbox {\vrule width.6pt  height5pt  
                      \kern5pt 
                      \vrule width.6pt  height5pt }
               \hrule height.6pt
               \hbox {\vrule width.6pt  height5pt  
                      \kern5pt 
                      \vrule width.6pt  height5pt }
               \hrule height.6pt}
                         }
              }
           \>\> }


\nref\SeibergI{N. Seiberg, Nucl. Phys. {\bf B435} (1995) 129.}
\nref\IntriligatorPouliot{K. Intriligator and P. Pouliot, Phys. Lett. 
  {\bf B353} (1995) 471.}
\nref\IntriligatorSeiberg{K. Intriligator and N. Seiberg, Nucl. Phys. 
  {\bf B444} (1995) 125.}
\nref\Kutasov{D. Kutasov, Phys. Lett. {\bf B351} (1995) 230.}
\nref\KutasovSchwimmer{D. Kutasov and A. Schwimmer, Phys. Lett. {\bf B354} 
  (1995) 315.}
\nref\Intriligator{K. Intriligator, Nucl. Phys. {\bf B448} (1995) 187.}
\nref\LS{R.G. Leigh and M.J. Strassler, Nucl. Phys. {\bf B447} (1995) 95.}
\nref\ILS{K. Intriligator, R.G. Leigh and M.J. Strassler, Nucl. Phys. 
  {\bf B456} (1995) 567.}
\nref\KSS{D. Kutasov, A. Schwimmer and N. Seiberg, Nucl. Phys. {\bf B459} 
  (1996) 455.}
\nref\Brodie{J. Brodie, Nucl. Phys. {\bf B478} (1996) 123.}
\nref\Ramond{P. Ramond, hep-th 9608077, unpublished.}
\nref\Distler{J. Distler and A. Karch, hep-th 9611088, unpublished.}
\nref\Csaki{C. Cs\`aki, M. Schmaltz, W. Skiba and J. Terning, hep-th 9701191, 
  unpublished.}
\nref\Bershadsky{M. Bershadsky, A. Johansen, T. Pantev, V. Sadov and C. Vafa, 
  hep-th 9612052, unpublished.}
\nref\Vafa{C. Vafa and B. Zweibach, hep-th 9701015, unpublished.}
\nref\Pouliot{P. Pouliot, Phys. Lett. {\bf B359} (1995) 108.}
\nref\PouliotStrasslerI{P. Pouliot and M. Strassler, Phys. Lett. {\bf B370} 
 (1996) 76.}
\nref\PouliotStrasslerII{P. Pouliot and M. Strassler, Phys. Lett. {\bf B375} 
 (1996) 175.}
\nref\Kawano{T. Kawano, Prog. Theor. Phys. {\bf 95} (1996) 963.}
\nref\Shifman{M.A. Shifman and A.I. Vainshtein, Nucl. Phys. {\bf B359} (1991) 
 571.}
\nref\Elashvili{A.G. Elashvili, Funk. Anal. Pril. {\bf 6} (1972) 51.}
\nref\Cho{P. Cho, hep-th 9701020, unpublished.}


\def\LongTitle#1#2#3{\nopagenumbers\abstractfont
\hsize=\hstitle\rightline{#1}
\vskip 0.5in\centerline{\titlefont #2} \centerline{\titlefont #3}
\abstractfont\vskip .3in\pageno=0}
 
\LongTitle{HUTP-97/A002}
{More on Chiral-Nonchiral Dual Pairs}{}

\centerline{
  Peter Cho\footnote{$^*$}{Research supported in part by the National Science 
Foundation under Grant \#PHY-9218167.}}
\centerline{Lyman Laboratory}
\centerline{Harvard University}
\centerline{Cambridge, MA  02138}

\vskip 0.3in
\centerline{\bf Abstract}
\bigskip

	Expanding upon earlier work of Pouliot and Strassler, we construct
chiral magnetic duals to nonchiral supersymmetric electric theories based upon 
$SO(7)$, $SO(8)$ and $SO(9)$ gauge groups with various numbers of vector and 
spinor matter superfields.  Anomalies are matched and gauge invariant operators 
are mapped within each dual pair.  Renormalization group flows along flat 
directions are also examined.  We find that confining phase quantum constraints 
in the electric theories are recovered from semiclassical equations of motion 
in their magnetic counterparts when the dual gauge groups are completely 
Higgsed.

\Date{2/97}

\newsec{Introduction}

	Although two years have passed since Seiberg's discovery of a dual to 
SUSY QCD \SeibergI, constructing weakly coupled descriptions of strongly 
interacting $\CN=1$ supersymmetric gauge theories remains at present more an 
art than a science.  Magnetic counterparts to other electric theories 
containing matter fields in only fundamental gauge group representations 
are qualitatively similar to those for SUSY QCD 
\refs{\IntriligatorPouliot,\IntriligatorSeiberg}.  Such models are 
sufficiently simple that duals to them can be deduced in the absence of any 
simplifying tree level superpotentials.  However, finding duals to theories 
with even slightly more complicated matter contents remains an outstanding 
challenge.  Several examples have been worked out in cases where operator 
chiral rings are truncated by classical equations of motion 
\refs{\Kutasov{--}\Brodie}.  A handful more have been identified 
where the magnetic and electric gauge groups are the same 
\refs{\Ramond{--}\Csaki}.  But to date, no basic algorithm has been developed 
which allows one to systematically establish long distance equivalence classes 
of different microscopic SUSY gauge theories.  Recent work of Vafa and 
collaborators offers the exciting prospect that $\CN=1$ duality fundamentally 
stems from T-duality within string theory \refs{\Bershadsky, \Vafa}.  
Hopefully, these string insights will lead to a deeper understanding of field 
theory duality. But for now, the subject resembles black magic.

	Among the duals which have been uncovered since Seiberg's seminal 
paper, one particular electric-magnetic pair found by Pouliot is especially
interesting \Pouliot.  The electric theory is based upon an $SO(7)$ gauge 
group with $\Nf$ spinor matter fields, while its magnetic counterpart has an 
$SU(\Nf-4)$ gauge group.
\foot{As we do not consider global gauge theory properties in this article, we 
do not distinguish between $SO(\Nc)$ and its covering group $Spin(\Nc)$.}
As the spinor of $SO(7)$ is real, the electric theory is nonchiral.  Yet the 
matter content in Pouliot's dual belongs to a manifestly chiral representation. 
Similar chiral-nonchiral dichotomies persist in generalizations of the $SO(7)$ 
model to $SO(8)$ and $SO(10)$ theories with one spinor and $\Nf$ vectors 
\refs{\PouliotStrasslerI{--}\Kawano}.  These models possess multiple 
flat directions which connect them to many other dual pairs.  They 
consequently provide valuable laboratories for probing various aspects of 
$\CN=1$ duality.

	In this article, we investigate further examples of chiral-nonchiral 
dual pairs.  More specifically, we construct and analyze duals to 
$SO(7)$, $SO(8)$ and $SO(9)$ models with $\Nf$ vectors and either one or two 
spinors.  We must note that all the results which we present here can be 
derived starting from the Pouliot-Strassler $SO(10)$ theory
\PouliotStrasslerII.  So none of the dualities that we will discuss are 
fundamentally new.  Nevertheless, the particular models that we focus upon 
and which have not previously been considered in the literature exhibit 
several interesting features.  They also provide useful reference points in the 
search for duals to even more complicated theories.  So they are worth 
studying in their own right.

	Our paper is organized as follows.  We describe in some detail 
our simplest dual to an $SO(7)$ model with one spinor and $\Nf$ vectors in 
section~2.  We then discuss qualitatively different aspects of duals to 
$SO(7)$, $SO(8)$ and $SO(9)$ theories with $\Nf$ vectors and respectively two 
8-dimensional spinors, one 8-dimensional spinor and its conjugate, and one 
16-dimensional spinor in sections 3, 4  and 5.  We pay particular attention to 
renormalization group flows among these models and to recovering 
electric theory quantum constraints from semiclassical magnetic equations of 
motion.  Finally, we tie together our results and close with some thoughts
on finding new duals in section~6. 

\newsec{$\pmb{SO(7)}$ with $\pmb{\Nf}$ vectors and one spinor}

	We begin our study by considering a supersymmetric gauge theory with 
symmetry group
\eqn\symgroupI{G = SO(7)_{\rm local} \times \bigl[ SU(\Nf) \times U(1)_\Y 
\times U(1)_\R \bigr]_{\rm global}}
and matter content
\eqn\matterI{\eqalign{
V^i_\mu & \sim \bigl( 7; \fund ; -1,1 - {4 \over \Nf} \bigr) \cr
Q^\A & \sim \bigl( 8; 1 ; \Nf, 0 \bigr). \cr}}
The hypercharge and R-charge assignments for the vector and spinor superfields 
are chosen so that both abelian factors in $G$ are anomaly free.  
Since the 7 and 8-dimensional irreps of $SO(7)$ are real, this theory 
is nonchiral.  It is also asymptotically free so long as its one-loop 
Wilsonian beta function coefficient
\foot{We adopt the $SO(\Nc)$ index values $K(\rm vector) = 2$, $K(\rm adjoint) 
= 2\Nc-4$, $K(\rm spinor) = 2^{\Nc-6 \over 2}$ for $\Nc$ even and 
$K(\rm spinor) =2^{\Nc-5 \over 2}$ for $\Nc$ odd.}
\eqn\betaI{b_0 = \half \bigl[ 3 K({\rm Adj}) -
\sum_{\rm{\buildrel matter \over {\scriptscriptstyle reps \>\rho}}} K(\rho)
\bigr] = 14 - \Nf}
is positive.  The full beta function which governs the running of the 
physical gauge coupling is then negative \Shifman.  The model's 
infrared dynamics are consequently nontrivial provided it contains $\Nf < 14$ 
vector flavors.

	Generic matter field expectation values break the $SO(7)$ gauge group 
according to the pattern \Elashvili
\eqn\patternI{SO(7) \> {\buildrel 8 \over \longrightarrow} \>
             G_2 \> {\buildrel 7 \over \longrightarrow} \> 
             SU(3) \> {\buildrel 7 \over \longrightarrow} \>
             SU(2) \> {\buildrel 7 \over \longrightarrow} \>
             1.}
This expression illustrates the hierarchy of gauge symmetries realized at 
progressively longer distance scales assuming that the spinor vev's magnitude 
is larger than the first vector's which in turn is larger than the second 
vector's and so on.  The symmetry breaking information displayed in \patternI\ 
allows one to readily count the gauge invariant operators that act as 
coordinates on the moduli space of degenerate vacua for small numbers of 
vector flavors \refs{\Pouliot,\Cho}.  In Table~1, we list the initial parton 
matter degrees of freedom, the unbroken color subgroup and the number of 
matter fields eaten by the superHiggs mechanism as a function of $\Nf$.
The remaining uneaten parton fields correspond to the independent hadrons in 
the low energy effective theory which label D-flat directions of the $SO(7)$ 
scalar potential. 

\midinsert
\parasize=1in

\begintable
$\quad \Nf \quad $ \| Parton DOF \| Unbroken Subgroup \| Eaten DOF
\| Hadrons \crthick
0 \| 8 \| $G_2$ \| $21-14=7$ \| 1 \nr
1 \| 15 \| $SU(3)$ \| $21-8=13$ \| 2 \nr
2 \| 22 \| $SU(2)$ \| $21-3=18$ \| 4 \nr
3 \| 29 \| 1 \| 21 \| 8 \nr
4 \| 36 \| 1 \| 21 \| 15 \endtable 
\bigskip
\centerline{Table 1: Number of independent hadron operators}
\endinsert

	The gauge invariant operators' flavor structures are tightly 
constrained by counting and symmetry considerations.  We first recall 
that the product of two $SO(7)$ spinors decomposes into a sum over rank-$n$ 
antisymmetric tensor irreps $[n]$ as
\eqn\sosevenspinorprod{8 \times 8 = [0]_\S + [1]_\A + [2]_\A + [3]_\S}
where the ``S'' and ``A'' subscripts indicate symmetry and antisymmetry under 
spinor exchange.   Since our model contains just one spinor flavor, we can 
only form hadrons involving spinor products in the symmetric $[0]$ 
and $[3]$ (mod 4) irreps.  Vector superfields can be contracted into these 
bispinor combinations using $SO(7)$ Gamma matrices $\Gamma_\mu$ and charge 
conjugation matrix $C$ as Clebsch-Gordan coefficients. We thus form the 
composite operators 
\eqn\eleccompositesI{\eqalign{
L &= Q^\T C Q \sim \bigl(1; 1; 2\Nf , 0 \bigr) \cr
M^{(ij)} &= (V^\T)^{i \mu} V^j_\mu 
   \sim \bigl(1; \sym; -2, 2 - {8 \over \Nf} \bigr) \cr
P^{[ijk]} &= {1 \over 3!} Q^\T \Vslash^{[i} \Vslash^j \Vslash^{k]} C Q 
   \sim \bigl(1; \antithree; 2 \Nf - 3, 3 - {12 \over \Nf} \bigr) \cr
R^{[ijkl]} &= {1 \over 4!} Q^\T \Vslash^{[i} \Vslash^j \Vslash^k \Vslash^{l]}
C Q \sim \bigl(1; \antifour; 2\Nf - 4, 4 - {16 \over \Nf} \bigr). \cr}}

	In order to determine whether these hadrons account for all massless 
fields in the $SO(7)$ model with four or fewer flavors, we tally their number 
as a function of $\Nf$ in Table~2.  For $\Nf \le 3$, the sum over the $L$, $M$ 
and $P$ degrees of freedom agrees with the required number of hadrons listed in 
Table~1.  On the other hand, the hadron count exceeds the needed number of 
composites by one when $\Nf=4$.  So a single constraint must exist among $L$, 
$M$, $P$ and $R$ in this case.  The precise quantum constraint relation is 
fixed by symmetries and the weakly coupled $\Lambda_4 \to 0$ classical limit. 
It appears in superpotential form as 
\eqn\WconstraintI{W_{\Nf=4} = X \bigl[ L^2 \det M + P_i M^{ij} P_j - R^2 
  - L \Lambda_4^{10}\bigr]}
where the prefactor $X$ represents a Lagrange multiplier field.  As a check 
on this expression, one can add a mass term for the spinor, integrate 
out heavy degrees of freedom and flow down to the $SO(7)$ theory 
with just $\Nf=4$ vectors.  The two separate branches in $\Nf=\Nc-3$ $SO(\Nc)$ 
theory with $W=0$ or $W \propto \Lambdab / \det M$ are then both properly 
recovered from $W_{\Nf=4}$ \refs{\IntriligatorSeiberg,\PouliotStrasslerI}.  

\topinsert
\parasize=1in
  
\begintable
$\quad \Nf \quad $ \| Hadrons \| $ \sp L \sp$ \| $\sp M \sp$ \| $\sp P \sp$
\| $\sp R \sp$ \| constraints \crthick
0 \| 1  \| 1 \|    \|   \|    \|    \cr
1 \| 2  \| 1 \| 1  \|   \|    \|    \cr
2 \| 4  \| 1 \| 3  \|   \|    \|    \cr
3 \| 8  \| 1 \| 6  \| 1 \|    \|    \cr
4 \| 15 \| 1 \| 10 \| 4 \| 1  \| -1 \endtable 
\bigskip
\centerline{Table 2: Hadron degree of freedom count}
\endinsert

	The quantum moduli space of degenerate vacua in the $\Nf=4$ model 
clearly differs from its classical progenitor as a result of the strong 
interaction scale $\Lambda_4$ appearing on the RHS of \WconstraintI.  Yet the 
origin $L=M=P=R=0$ lies on both.  Since the global $SU(\Nf) \times U(1)\Y 
\times U(1)_\R$ symmetry group remains unbroken at this point, it is 
instructive to compare 't~Hooft anomalies in the microscopic and macroscopic 
theories.  We find that the $SU(\Nf)^3$, $SU(\Nf)^2 U(1)_\Y$, 
$SU(\Nf)^2 U(1)_\R$, $U(1)_\Y$, $U(1)_\Y^3$, $U(1)_\R$, $U(1)_\R^3$, 
$U(1)_\Y^2 U(1)_\R$ and $U(1)_\R^2 U(1)_\Y$ anomalies all match provided we 
include a field $X \sim (1; 1; -8, 2)$ into the low energy spectrum which has 
the same quantum numbers as the $\Nf=4$ Lagrange multiplier.  This nontrivial 
agreement among nine different 't~Hooft anomalies strongly suggests that the 
colorless composites in \eleccompositesI\ represent the only light 
degrees of freedom in the low energy theory.  It is natural to interpret this 
finding as evidence for confinement in the $\Nf=4$ $SO(7)$ model.  On the 
other hand, the parton and hadron level global anomalies do not match when 
$\Nf=5$, and the disagreement cannot be eliminated via inclusion of additional 
color-singlet fields without disrupting the $\Nf=4$ results.  So we conclude 
that the $SO(7)$ model ceases to confine at this juncture.  

	The vacuum structure of the $\Nf=5$ model is more simply understood in 
terms of a weakly coupled magnetic dual to the $SO(7)$ electric theory.  The 
dual description has symmetry group 
\eqn\maggroupI{\Gdual = SU(\Nf-3)_{\rm local} \times \bigl[ SU(\Nf) 
\times U(1)_\Y \times U(1)_\R \bigr]_{\rm global},}
superfield matter content
\eqn\magmatterI{\eqalign{
q^\a_i &\sim \bigl( \fund; \antifund; {2\Nf-3 \over \Nf-3}, {3 \over \Nf} 
  {\Nf-4 \over \Nf-3} \bigr) \cr
\qp^\a &\sim \bigl( \fund; 1; {\Nf \over \Nf-3}, {\Nf-4 \over \Nf-3} \bigr) 
  \cr
s_{\a\b} &\sim \bigl( \symbar; 1; -{2 \Nf \over \Nf-3}, {2 \over \Nf-3} 
  \bigr) \cr
\ell &\sim (1; 1; 2\Nf, 0 \bigr) \cr
m^{(ij)} &\sim \bigl(1; \sym; -2, 2 - {8 \over \Nf} \bigr) \cr}}
and tree level superpotential 
\eqn\WmagI{
\Wmag = \mu_1^{-2} m^{(ij)} q^\a_i s_{\a\b} q^\b_j 
  - \qp^\a s_{\a\b} \qp^\b - \mu_2^{4-\Nf} \ell \det s.}
Several points about this magnetic theory should be noted.  Firstly, its 
color gauge group and abelian hypercharge and R-charge global symmetries 
are nonanomalous like their electric counterparts.  Moreover, 
all nine global `t Hooft anomalies match between the electric and magnetic 
sides for $\Nf \ge 4$.  This highly nontrivial anomaly 
agreement provides an important first check on the duality.  Secondly, the 
magnetic Wilsonian beta function coefficient $\widetilde{b_0} = 2 \Nf - 9$ is 
positive for $\Nf \ge 5$.  The dual theory is consequently asymptotically 
free and does not possess a free magnetic phase.  Instead, it exists at the
origin of moduli space in a nonabelian Coulomb phase for $5 \le \Nf < 14$ 
flavors.  Thirdly, the dual global symmetry group starts as 
$\Gdual_{\rm global} = U(\Nf+1)_{q+q'} 
\times U(1)_s \times U[1+\half\Nf(\Nf+1)]_{\ell+m}$ in the absence of any 
tree level superpotential.  The interaction terms in \WmagI\ 
break $\Gdual_{\rm global}$ down to $SU(\Nf) \times U(1)_\Y \times U(1)_\R$.  
The microscopic electric and magnetic theories therefore share the same 
global symmetry group at short as well as long distance scales.  Fourthly,
we treat all colored elementary fields in the magnetic theory as canonically 
normalized.  On the other hand, we take the ultraviolet mass dimensions for 
the color-singlets to equal those of their electric analogues.  In order 
for the magnetic superpotential to have dimension three, its nonrenormalizable 
interaction terms must be accompanied by appropriate inverse powers of some 
scales $\mu_1$ and $\mu_2$.  For simplicity, we set these scales equal to 
unity from here on.  Finally, the magnetic theory's matter content belongs to 
a manifestly chiral representation of $\Gdual$.  So this dual to the $SO(7)$ 
model represents another example of a chiral-nonchiral pair.

	In order to check the dual, we look for maps between gauge invariant 
composites in the electric and magnetic theories.  The $L$ and $M^{(ij)}$ 
hadrons in \eleccompositesI\ are obviously identified with the elementary 
$\ell$ and $m^{(ij)}$ fields in \magmatterI.  $P^{[ijk]}$ and $R^{[ijkl]}$ are 
matched onto the dual quark combinations
\eqn\magbaryonsI{\eqalign{
p^{[i_1 i_2 i_3]} &= \e^{i_1 \cdots i_\Nf} \e_{\a_1 \cdots \a_{\Nf-3}} 
 \sp q^{\a_1}_{i_4} \cdots q^{\a_{\Nf-3}}_{i_\Nf} \cr
r^{[i_1 \cdots i_4]} &= \e^{i_1 \cdots i_\Nf} \e_{\a_1 \cdots \a_{\Nf-3}} 
  \sp q^{\a_1}_{i_5} \cdots q^{\a_{\Nf-4}}_{i_\Nf} \qp^{\a_{\Nf-3}}. \cr}}
These magnetic baryons possess the same global quantum numbers as their 
electric theory counterparts.  Additional gauge invariant operators beyond 
those relevant for the $SO(7)$ model's confining phase can also be mapped.
For instance, the electric baryons 
\eqn\elecbaryonsI{\eqalign{
B_0^{[i_1 \cdots i_7]} &= \e^{\mu_1 \cdots \mu_7} V^{i_1}_{\mu_1} \cdots 
  V^{i_7}_{\mu_7} \cr
B_1^{a[i_1 \cdots i_5]} &= \e^{\mu_1 \cdots \mu_7} V^{i_1}_{\mu_1} \cdots
V^{i_5}_{\mu_5} W^a_{\mu_6 \mu_7} \cr
B_2^{[i_1 i_2 i_3]} &= \e^{\mu_1 \cdots \mu_7} V^{i_1}_{\mu_1} V^{i_2}_{\mu_2}
V^{i_3}_{\mu_3} W^a_{\mu_4 \mu_5} W^a_{\mu_6 \mu_7} \cr}}
are identified with magnetic antibaryons built from effective dual 
antiquarks $(\qbar_{\rm eff})_{\a i} \equiv s_{\a\b} q^\a_i$:
\eqn\magantibaryonsI{\eqalign{
b_0^{[i_1 \cdots i_7]} &= \e^{i_1 \cdots i_\Nf} \e^{\a_1 \cdots \a_{\Nf-3}} 
  \sp (sq)_{\a_1 i_8} \cdots (sq)_{\a_{\Nf-7} i_\Nf} (s \wtilde)_{\a_{\Nf-6}
\a_{\Nf-5}} (s \wtilde)_{\a_{\Nf-4} \a_{\Nf-3}} \cr
b_1^{a[i_1 \cdots i_5]} &= \e^{i_1 \cdots i_\Nf} \e^{\a_1 \cdots \a_{\Nf-3}} 
  \sp (sq)_{\a_1 i_6} \cdots (sq)_{\a_{\Nf-5} i_\Nf} (s \wtilde)_{\a_{\Nf-4}
\a_{\Nf-3}} \cr
b_2^{[i_1 i_2 i_3]} &= \e^{i_1 \cdots i_\Nf} \e^{\a_1 \cdots \a_{\Nf-3}} 
  \sp (sq)_{\a_1 i_4} \cdots (sq)_{\a_{\Nf-3} i_\Nf}. \cr }}
The $B_1^a$ superfield has a simple interpretation in the $\Nf \ge 5$ 
electric theory when viewed along flat directions where the $SO(7)$ gauge group 
is Higgsed down to $SO(2) \simeq U(1)$: it projects out the abelian 
Coulomb phase photon from the gluon field strength tensor 
\SeibergI.  Similarly, $B_2$ becomes a second glueball in addition to 
$S=W^a_{\mu\nu} W^{\mu\nu a}$ along flat directions where $SO(7)$ breaks to 
$SO(4) \simeq SU(2) \times SU(2)$.  

	To further test the duality, we investigate deformations of the 
electric and magnetic theories which induce renormalization group flows to 
known dual pairs.  We first consider Higgsing $SO(7)$ along the spinor flat 
direction.  When $Q^\A$ acquires a nonvanishing expectation value, 
the electric theory's symmetry group breaks down to 
\eqn\symsubgroupI{H = \bigl[G_2 \bigr]_{\rm local} \times 
\bigl[ SU(\Nf) \times U(1)_\R \bigr]_{\rm global},}
and its 7-dimensional vectors become fundamentals under the $G_2$ subgroup:
\eqn\submatterI{F^i_\a \sim \bigl(7 ; \fund; 1 - {4 \over \Nf} \bigr).}
The global hypercharge in \symgroupI\ is broken by the spinor vev, but the 
$U(1)_\R$ factor remains unaffected since $Q^\A$ carries zero R-charge.
On the magnetic side, we freeze $\ell$ at its induced expectation 
value and drop the hypercharge assignments for all fields.  The dual theory 
consequently reduces to 
\eqn\magsubgroupI{\Hdual = SU(\Nf-3)_{\rm local} \times \bigl[ SU(\Nf) 
\times U(1)_\R \bigr]_{\rm global}}
with matter content
\eqn\GtwomatterI{\eqalign{
q^\a_i &\sim \bigl( \fund; \antifund; {3 \over \Nf} 
  {\Nf-4 \over \Nf-3} \bigr) \cr
\qp^\a &\sim \bigl( \fund; 1; {\Nf-4  \over \Nf-3} \bigr) \cr
s_{\a\b} &\sim \bigl( \symbar; 1; {2 \over \Nf-3} \bigr) \cr
m^{(ij)} &\sim \bigl(1; \sym; 2 - {8 \over \Nf} \bigr) \cr}}
and tree level superpotential 
\eqn\WGtwo{
\Wmag = m^{(ij)} q^\a_i s_{\a\b} q^\b_j - \qp^\a s_{\a\b} \qp^\b - \det s.}
This deformation reproduces Pouliot's dual to $G_2$ theory with $\Nf$ 
fundamentals \Pouliot.

	We next consider adding a mass term for the spinor field.  The $SO(7)$ 
model then becomes 
\eqn\deformedsymgroupI{G = SO(7)_{\rm local} \times \bigl[ SU(\Nf) 
\times U(1)_{\R'} \bigr]_{\rm global}}
with 
\eqn\deformedmatterI{
V^i_\mu \sim \bigl( 7; \fund ; 1 - {5 \over \Nf} \bigr).}
In the magnetic theory, we add $\Wtree = \mu \ell$ to the superpotential 
in \WmagI\ which causes $s_{\a\b}$ to condense.  Its expectation value 
$\vev{s_{\a\b}}$ can always be rotated into diagonal form via an $SU(\Nf-3)$ 
color transformation.  But from the $\ell$ field's equation of motion, we 
learn that $\det s = \mu$ is nonvanishing.  We may consequently perform a 
nonsingular complexified $SU(\Nf-3)$ rotation and set each nonzero 
eigenvalue of $\vev{s_{\a\b}}$ equal to unity.  The $s$ field's identity 
matrix vev then breaks the magnetic gauge group $SU(\Nf-3)$ down to 
$SO(\Nf-3)$ and yields a mass term for the $q'$ dual quark.  After 
integrating out $q'$, we find that the deformed magnetic theory reduces 
precisely to Seiberg's dual for an $SO(7)$ theory with $\Nf$ vectors 
\refs{\SeibergI,\PouliotStrasslerI}.

	The most interesting duality result emerges when we give mass to all 
but four flavors in the electric theory and thereby completely Higgs the 
magnetic gauge group. Instanton effects generate a new superpotential 
term beyond those already present in \WmagI:
\eqn\WmagIfour{\Wmag \> {\buildrel \Nf \to 4 \over \longrightarrow} \> 
 s \bigl[ \ell^2 \det m + q_i m^{(ij)} q_j - q' q' - \ell \bigr].}
Each elementary dual field in this expression can be identified with 
a corresponding composite hadron in the $\Nf=4$ $SO(7)$ theory on the basis 
of its charge assignments:
\eqn\fieldmapsI{\eqalign{
q_i & \sim (1; \antifund; 5,0) \sim P_i \cr
q' & \sim (1; 1; 4,0) \sim R \cr
s & \sim (1;1;-8,2) \sim X \cr
\ell & \sim (1; 1; 8, -8) \sim L \cr
m^{(ij)} & \sim (1; \sym; -2,2) \sim M^{(ij)}. \cr}}
It is important to note that $s$ has exactly the right quantum numbers to be 
mapped onto the $X$ Lagrange multiplier.  After replacing the magnetic fields 
in \WmagIfour\ by their electric counterparts and restoring dimensionful 
scales, we reproduce the confining phase $\Nf=4$ superpotential in 
\WconstraintI.  The quantum constraint among hadrons in the electric theory is 
thus recovered from semiclassical equations of motion in the magnetic theory.  
So we see duality at work relating complicated nonperturbative dynamics in one 
description to weakly coupled phenomena in the other.

\newsec{$\pmb{SO(7)}$ with $\pmb{\Nf}$ vectors and two spinors}

	In this section, we generalize our preceding discussion by adding 
a second spinor into the $SO(7)$ model.  The modified electric theory now has 
symmetry group 
\eqn\symgroupII{G = SO(7)_{\rm local} \times \bigl[ SU(\Nf) \times SU(2) 
\times U(1)_\Y \times U(1)_\R \bigr]_{\rm global},}
superfield matter content
\eqn\matterII{\eqalign{
V^i_\mu & \sim \bigl( 7; \fund ,1 ; -2, 1 - {5 \over \Nf} \bigr) \cr
Q^\A_\I  & \sim \bigl( 8; 1,2 ; \Nf, 1 \bigr) \cr}}
and Wilsonian beta function coefficient $b_0 = 13 - \Nf$.  The global $SU(2)$ 
factor which rotates the two spinor fields represents a qualitatively new 
feature in this model compared to the previous one.  We will shortly see how 
it is implemented in the magnetic dual.

	Given the pattern of symmetry breaking realized at generic points in 
moduli space~\Elashvili
\eqn\patternII{SO(7) \> {\buildrel 8 \over \longrightarrow} \>
               G_2 \> {\buildrel 8 \over \longrightarrow} \> 
               SU(3) \> {\buildrel 7 \over \longrightarrow} \>
               SU(2) \> {\buildrel 7 \over \longrightarrow} \>
               1,}
it is straightforward to determine the number of independent gauge invariant 
operators needed to label flat directions in the second $SO(7)$ model as a 
function of $\Nf$.  Using the same sorts of counting arguments as in 
section~2, we find that the hadrons
\eqn\eleccompositesII{\eqalign{
L_\xrm  &= Q^\T_\I (\s_2 \s_\xrm)_{\I\J} C Q_\J 
	\sim \bigl(1; 1, 3; 2\Nf, 2 \bigr) \cr
M^{(ij)} &= (V^\T)^{i \mu} V^j_\mu 
	\sim \bigl(1; \sym, 1; -4, 2 - {10 \over \Nf} \bigr) \cr
N^i &= Q^\T_\I (\s_2)_{\I\J} \Vslash^i C Q_\J 
	\sim \bigl(1; \fund,1; 2 \Nf - 2, 3 - {5 \over \Nf} \bigr) \cr
O^{[ij]} &= {1 \over 2!} Q^\T_\I (\s_2)_{\I\J} \Vslash^{[i} \Vslash^{j]} C Q_\J
	\sim \bigl(1; \anti; 2\Nf - 4, 4 - {10 \over \Nf} \bigr) \cr
P^{[ijk]}_\xrm &= {1 \over 3!} Q^\T_\I (\s_2 \s_\xrm)_{\I\J} \Vslash^{[i} 
  \Vslash^j \Vslash^{k]} C Q_\J 
	\sim \bigl(1; \antithree, 3; 2\Nf - 6, 5 - {15 \over \Nf} \bigr)\cr }}
account for all massless degrees of freedom when $\Nf \le 2$.  In these 
operator definitions, $SO(7)$ color, $SU(\Nf)$ vector and $SU(2)$ spinor 
indices are respectively denoted by Greek, small Latin and large Latin 
letters.  We also note that the Pauli matrix factors $\s_2 \s_\X$ and $\s_2$ 
combine together spinors into symmetric and antisymmetric products in accord 
with \sosevenspinorprod.  

	The hadron count exceeds the required number of composites by two 
when $\Nf=3$.  So two separate constraints must exist among $L$, $M$, 
$N$, $O$ and $P$ in this case.  These quantum relations are fixed by 
symmetry considerations and the $\Lambda_3 \to 0$ classical limit: 
\eqn\WconstraintII{\eqalign{
W_{\Nf=3} &= X \bigl[ L_\xrm L_\xrm \det M - \half \e_{i_1 i_2 i_3} 
\e_{j_1 j_2 j_3} M^{i_1 j_1} M^{i_2 j_2} N^{i_3} N^{j_3} + O_i M^{ij} O_j 
- P_\xrm P_\xrm - \Lambda_3^{10} \bigr] \cr
& \qquad + Y\bigl[ N^i O_i - L_\xrm P_\xrm \bigr]. \cr}}
The first constraint prohibits the point $L=M=N=O=P=0$ from lying on the 
quantum moduli space.  As a result, matching global `t~Hooft anomalies at 
the origin where all global symmetries are unbroken might seem to make little 
sense.  Nevertheless, we find that the parton and hadron level $SU(\Nf)^3$, 
$SU(\Nf)^2 U(1)_\Y$, $SU(\Nf)^2 U(1)_\R$, $SU(2)^2 U(1)_\Y$, $SU(2)^2 U(1)_\R$, 
$U(1)_\Y$, $U(1)_\Y^3$, $U(1)_\R$, $U(1)_\R^3$, $U(1)_\Y^2 U(1)_\R$ and 
$U(1)_\R^2 U(1)_\Y$ anomalies all agree when $\Nf=3$ provided we include 
the Lagrange multipliers $X \sim (1;1,1; 0,2)$ and $Y \sim (1; 1,1; -6,0)$
into the low energy spectrum.  As we shall see, the magnetic dual provides an
explanation for this surprising electric theory result.

	The dual to the second $SO(7)$ model has symmetry group
\eqn\maggroupII{\Gdual = SU(\Nf-2)_{\rm local} \times \bigl[ SU(\Nf) 
\times SO(3) \times U(1)_\Y \times U(1)_\R \bigr]_{\rm global},}
matter content
\eqn\magmatterII{\eqalign{
q^\a_i &\sim \bigl( \fund ; \antifund,1; 2, {4 \Nf-10 \over \Nf(\Nf-2)} 
  \bigr) \cr
\qp^\a_\xrm &\sim \bigl( \fund; 1,3; 0, {\Nf-3 \over \Nf-2} \bigr) 
  \cr
\qbar_\a &\sim \bigl( \antifund; 1,1; -2\Nf, -{\Nf-3 \over \Nf-2} \bigr) \cr
s_{\a\b} &\sim \bigl( \symbar; 1,1; 0, {2 \over \Nf-2} \bigr) \cr
\ell_\xrm &\sim (1; 1,3; 2\Nf, 2 \bigr) \cr
m^{(ij)} &\sim \bigl(1; \sym,1; -4, 2 - {10 \over \Nf} \bigr) \cr
n^i &\sim \bigl(1; \fund,1; 2\Nf-2, 3 - {5 \over \Nf} \bigr) \cr}}
and tree level superpotential
\eqn\WmagII{
\Wmag = \mu_1^{-2} n^i q^\a_i \qbar_\a 
  - \mu_2^{-1} \ell_\xrm \qp^\a_\xrm \qbar_\a 
  + \mu_3^{-2} m^{(ij)} q^\a_i s_{\a\b} q^\b_j 
  - \qp^\a_\xrm s_{\a\b} \qp^\b_\xrm 
  - \mu_4^{5-\Nf} \det s.}
Various points about this dual description should be noted.  Firstly, 
the magnetic gauge group and global hypercharge and R-charge symmetries are 
nonanomalous, and all global 't~Hooft anomalies match between the electric and 
magnetic theories.  Secondly, the magnetic beta function coefficient 
$\widetilde{b_0} = 2 \Nf - 8$ is positive for $\Nf \ge 5$.  When $\Nf=4$, the 
last determinant term in $\Wmag$ reduces to a quadratic mass term for 
$s_{\a\b}$.  After removing this field from the low energy effective theory, 
we find $\widetilde{b_0} = 2$.  So the magnetic theory is asymptotically free 
and exists at the moduli space origin in a nonabelian Coulomb phase for 
$4 \le \Nf < 13$.  Finally, the magnetic superpotential interactions break the 
initial $U(\Nf+3)_{q+q'} \times U(1)_\qbar \times U(1)_s \times 
U\bigl[3+\Nf+\half \Nf(\Nf+1)\bigr]_{\ell+m+n}$ global symmetry down to 
$\Gdual_{\rm global} = SU(\Nf) \times SO(3) \times U(1) \times U(1)$.  We 
again set the various dimensionful scales that appear in $\Wmag$ to unity for 
simplicity.  

	Mappings between electric and magnetic gauge invariant composites 
in this second $SO(7)$ model are straightforward generalizations of those in 
the first.  The $L_\xrm$, $M^{(ij)}$ and $N^i$ hadrons in 
\eleccompositesII\ are identified with their $\ell_\xrm$, $m^{(ij)}$ and 
$n^i$ counterparts in \magmatterII, while $O^{[ij]}$ and $P^{[ijk]}_\xrm$ are 
matched onto 
\eqn\magbaryonsII{\eqalign{
o^{[i_1 i_2]} &= \e^{i_1 \cdots i_\Nf} \e_{\a_1 \cdots \a_{\Nf-2}} 
 \sp q^{\a_1}_{i_3} \cdots q^{\a_{\Nf-2}}_{i_\Nf} \cr
p^{[i_1 i_2 i_3]}_\xrm &= \e^{i_1 \cdots i_\Nf} \e_{\a_1 \cdots \a_{\Nf-2}} 
  \sp q^{\a_1}_{i_4} \cdots q^{\a_{\Nf-3}}_{i_\Nf} \qp^{\a_{\Nf-2}}_\xrm. \cr}}
The electric baryons 
\eqn\elecbaryonsII{\eqalign{
B_0^{[i_1 \cdots i_7]} &= \e^{\mu_1 \cdots \mu_7} V^{i_1}_{\mu_1} \cdots 
  V^{i_7}_{\mu_7} \cr
B_1^{a[i_1 \cdots i_5]} &= \e^{\mu_1 \cdots \mu_7} V^{i_1}_{\mu_1} \cdots
  V^{i_5}_{\mu_5} W^a_{\mu_6 \mu_7} \cr
B_2^{[i_1 i_2 i_3]} &= \e^{\mu_1 \cdots \mu_7} V^{i_1}_{\mu_1} V^{i_2}_{\mu_2}
V^{i_3}_{\mu_3} W^a_{\mu_4 \mu_5} W^a_{\mu_6 \mu_7} \cr}}
are identified with the magnetic composites 
\eqn\magantibaryonsII{\eqalign{
b_0^{[i_1 \cdots i_7]} &= \e^{i_1 \cdots i_\Nf} \e^{\a_1 \cdots \a_{\Nf-2}} 
  \sp (sq)_{\a_1 i_8} \cdots (sq)_{\a_{\Nf-7} i_\Nf} (s \wtilde)_{\a_{\Nf-6}
\a_{\Nf-5}} (s \wtilde)_{\a_{\Nf-4} \a_{\Nf-3}} \qbar_{\a_{\Nf-2}} \cr
b_1^{a[i_1 \cdots i_5]} &= \e^{i_1 \cdots i_\Nf} \e^{\a_1 \cdots \a_{\Nf-2}} 
  \sp (sq)_{\a_1 i_6} \cdots (sq)_{\a_{\Nf-5} i_\Nf} (s \wtilde)_{\a_{\Nf-4}
\a_{\Nf-3}} \qbar_{\a_{\Nf-2}} \cr
b_2^{[i_1 i_2 i_3]} &= \e^{i_1 \cdots i_\Nf} \e^{\a_1 \cdots \a_{\Nf-2}} 
  \sp (sq)_{\a_1 i_4} \cdots (sq)_{\a_{\Nf-3} i_\Nf} \qbar_{\a_{\Nf-2}}. \cr}}
We can also form more exotic combinations of vectors, spinors and 
gluons
\eqn\elecexoticsII{\eqalign{
C_0^{[i_1 \cdots i_6]} &= \e^{\mu_1 \cdots \mu_7} V^{i_1}_{\mu_1} \cdots 
  V^{i_6}_{\mu_6} Q^\T_\I (\s_2)_{\I\J} \Gamma_{\mu_7} C Q_\J \cr
C_1^{a [i_1 \cdots i_4]} &= \e^{\mu_1 \cdots \mu_7} V^{i_1}_{\mu_1} \cdots 
  V^{i_4}_{\mu_4} W^a_{\mu_5 \mu_6} Q^\T_\I (\s_2)_{\I\J} \Gamma_{\mu_7} C Q_\J 
  \cr
C_2^{[i_1 i_2]} &= \e^{\mu_1 \cdots \mu_7} V^{i_1}_{\mu_1} 
  V^{i_2}_{\mu_2} W^a_{\mu_3 \mu_4} W^a_{\mu_5 \mu_6} 
  Q^\T_\I (\s_2)_{\I\J} \Gamma_{\mu_7} C Q_\J \cr}}
which map onto the dual antibaryons
\eqn\magexoticsII{\eqalign{
c_0^{[i_1 \cdots i_6]} &= \e^{i_1 \cdots i_\Nf} \e^{\a_1 \cdots \a_{\Nf-2}} 
  \sp (sq)_{\a_1 i_7} \cdots (sq)_{\a_{\Nf-6} i_\Nf} (s \wtilde)_{\a_{\Nf-5}
\a_{\Nf-4}} (s \wtilde)_{\a_{\Nf-3} \a_{\Nf-2}} \cr
c_1^{a[i_1 \cdots i_4]} &= \e^{i_1 \cdots i_\Nf} \e^{\a_1 \cdots \a_{\Nf-2}} 
  \sp (sq)_{\a_1 i_5} \cdots (sq)_{\a_{\Nf-4} i_\Nf} (s \wtilde)_{\a_{\Nf-3}
\a_{\Nf-2}} \cr
c_2^{[i_1 i_2]} &= \e^{i_1 \cdots i_\Nf} \e^{\a_1 \cdots \a_{\Nf-2}} 
  \sp (sq)_{\a_1 i_3} \cdots (sq)_{\a_{\Nf-2} i_\Nf}. \cr }}
The consistency of all these operator identifications provides strong support 
for the duality hypothesis.

	The long distance equivalence between the microscopic $SO(7)$ and 
$SU(\Nf-2)$ theories can be probed via deformations along various flat 
directions.  Rather than repeat the same checks as described in 
section~2, we focus instead upon recovering our first model from the second.  
We start by adding the spinor mass term $\Wtree = \mu Q_2^\T C Q_2 \simeq 
- \half \mu (\ell_1 + i \ell_2)$ to both the 
electric and magnetic theories.  In the former, we simply integrate out $Q_2$ 
and flow down to the $SO(7)$ model with just one spinor.  In the latter, 
we eliminate the fields $\ell_1$, $\ell_2$ and $n^i$ which grow heavy since 
their electric counterparts contain the second spinor.  Their equations 
of motion force the condensates 
\eqn\condensatesII{\eqalign{
\vev{{q'}^\a_1} &= \bigl(0, \cdots, 0, \sqrt{\mu \over 2} \bigr) \cr
\vev{{q'}^\a_2} &= \bigl(0, \cdots, 0, i \sqrt{\mu \over 2} \bigr) \cr
\vev{\qbar_\a} &= \bigl(0, \cdots, 0, \sqrt{\mu \over 2} \bigr) \cr
\vev{q^{\Nf-2}_i} &= 0 \cr}}
to develop which break the $SU(\Nf-2)$ gauge group down to $SU(\Nf-3)$.  
After these vevs are substituted back into $\Wmag$ and the fields $\ell_3$ and 
${q'}^{\Nf-2}_3$ that acquire induced masses are integrated out, the 
superpotential reduces to 
\eqn\WmagIIreduced{\Wmag \to m^{(ij)} \qhat^\a_i \shat_{\a\b} \qhat^\b_j 
- \qhat^{'\a}_3 \shat_{\a\b} \qhat^{'\b}_3 
- \bigl( s_{\scriptscriptstyle{\Nf-2 \> \Nf-2}} \bigr) \det \shat}
where all hatted objects transform under $SU(\Nf-3)$.  This expression 
reproduces the superpotential in \WmagI\ when all hats are dropped and 
the fields $\qhat^{'\a}_3 \to \qp^a$ and $s_{\scriptscriptstyle{
\Nf-2 \> \Nf-2}} \to \ell$ are renamed.   We thus recover the dual to our 
first $SO(7)$ model.

	It is once again instructive to investigate the changes that arise in 
the dual pair when the number of vector flavors in the electric theory is 
reduced down to $\Nf=3$ and the magnetic gauge group becomes totally Higgsed.
In this case, the superpotential in \WmagII\ reduces to 
\eqn\WmagIIthree{\eqalign{
\Wmag \> {\buildrel \Nf \to 3 \over \longrightarrow} \> &
 s \bigl[ \ell_\xrm \ell_\xrm \det m - \half \e_{i_1 i_2 i_3} \e_{j_1 j_2 
j_3} m^{i_1 j_1} m^{i_2 j_2} n^{i_3 j_3} + q_i m^{(ij)} q_j - q'_\xrm q'_\xrm
 - 1 \bigr] \cr
& \quad + \qbar \bigl[ n^i q_i - \ell_\xrm q'_\xrm \bigr]. \cr}}
We have added by hand the first two terms on the RHS even though we have not 
identified their physical origin.  Consistency with confining phase 
results requires that these terms be generated.  After identifying all the 
dual fields in \WmagIIthree\ with confined electric composites on the basis 
of their charge assignments
\eqn\fieldmapsII{\eqalign{
q_i & \sim (1; \antifund,1; 2, {2 \over 3} ) \sim O_i \cr
q'_\xrm  & \sim (1; 1,3; 0,0) \sim P_\xrm \cr
\qbar & \sim (1; 1,1; -6,0) \sim Y \cr
s & \sim (1;1,1;0,2) \sim X \cr
\ell_\xrm & \sim (1; 1,3; 6, 2) \sim L_\xrm  \cr
m^{(ij)} & \sim (1; \sym,1; -4,{4 \over 3}) \sim M^{(ij)} \cr
n^i & \sim (1; \fund,1; 4, \third) \sim N^i, \cr}}
we recover the $\Nf=3$ superpotential in \WconstraintII.  
It is important to note that $s$ and $\qbar$ have exactly the same quantum 
numbers as the $X$ and $Y$ Lagrange multipliers.  So we now understand the 
origin of these multiplier fields in the low energy effective theory.  We 
also observe that the general agreement between electric and magnetic global 
anomalies which continues to hold even when the dual gauge group is completely 
Higgsed in conjunction with the field identifications in \fieldmapsII\ ensures 
parton and hadron level anomaly matching in the $\Nf=3$ model.  So as 
advertised, duality sheds light upon the `t~Hooft anomaly agreement. 

\newsec{$\pmb{SO(8)}$ with $\pmb{\Nf}$ vectors, one spinor and one conjugate 
spinor}

	The electric theory in the third chiral-nonchiral dual pair which 
we shall investigate has symmetry group 
\eqn\symgroupIII{G = SO(8)_{\rm local} \times \bigl[ SU(\Nf) \times U(1)_\Y
\times U(1)_\B \times U(1)_\R \bigr]_{\rm global},} 
matter superfields
\eqn\matterIII{\eqalign{
V^i_\mu & \sim \bigl( 8_\V ; \fund; -2,0,1-{6 \over \Nf} \bigr) \cr
Q^\A_\S & \sim \bigl( 8_\S ; 1; \Nf, 1, 1 \bigr) \cr 
Q^{\dot \A}_\C & \sim \bigl( 8_\C ; 1; \Nf, -1, 1 \bigr) \cr}}
and Wilsonian beta function coefficient $b_0 = 16 - \Nf$.  The discrete 
triality transformation that shuffles the vector, spinor and conjugate spinor 
representations among each other relates this model to two others with 
different 8-dimensional irrep assignments.  As the structure of the $SO(8)$ 
theory and its dual are similar to the preceding $SO(7)$ models, we will only 
briefly sketch their basic outlines in this section.

	Simple counting arguments 
demonstrate that the composite operators
\eqn\eleccompositesIII{\eqalign{
L_\S &= Q_\S^\T C Q_\S
	\sim \bigl(1; 1; 2\Nf , 2,2 \bigr) \cr
L_\C &= Q_\C^\T C Q_\C 
	\sim \bigl(1; 1; 2\Nf , -2,2 \bigr) \cr
M^{(ij)} &= (V^\T)^{i \mu} V^j_\mu 
	\sim \bigl(1; \sym; -4,0, 2-{12 \over \Nf} \bigr) \cr
N^i &= Q_\S^\T \Vslash^i C Q_\C
	\sim \bigl(1; \antithree; 2 \Nf - 2,0,3 - {6\over \Nf} \bigr) \cr
P^{[ijk]} &= {1 \over 3!} Q_\S^\T \Vslash^{[i} \Vslash^j \Vslash^{k]} C Q_\C 
	\sim \bigl(1; \antithree; 2\Nf - 6, 0 , 5 - {18 \over \Nf} \bigr) \cr 
R_\S^{[ijkl]} &= {1 \over 4!} Q_\S^\T \Vslash^{[i} \Vslash^j \Vslash^k 
\Vslash^{l]} C Q_\S  
	\sim \bigl(1; \antifour; 2\Nf - 8, 2, 6 - {24 \over \Nf} \bigr) \cr 
R_\C^{[ijkl]} &= {1 \over 4!} Q_\C^\T \Vslash^{[i} \Vslash^j \Vslash^k 
\Vslash^{l]} C Q_\C 
	\sim \bigl(1; \antifour; 2\Nf - 8,-2, 6 - { 24 \over \Nf} \bigr) \cr }}
account for all massless degrees of freedom in the $SO(8)$ model with $\Nf 
\le 3$ vector flavors.  In these operator definitions, we treat the spinor 
and conjugate spinor fields as the projections $Q_\S = P_+ Q$ and 
$Q_\C = P_- Q$ where $Q$ denotes a 16-dimensional spinor of $SO(9)$ and 
$P_\pm = \half (1 \pm \Gamma_9)$.  The $16 \times 16$ Gamma matrices hidden
inside $\Vslash^i = \sum_{i=1}^8 V^i_\mu \Gamma^\mu$ along with the charge 
conjugation matrix $C$ therefore originate from the Clifford algebra for 
$SO(9)$.

	The same degree of freedom counting indicates that the hadrons in 
\eleccompositesIII\ are not all independent when $\Nf=4$.  They are instead 
restricted by the two quantum constraints displayed below in superpotential 
form:
\eqn\WconstraintIII{\eqalign{W_{\Nf=4} &= 
X \bigl[ L_\S L_\C \det M - {1 \over 3!} \e_{i_1 i_2 i_3 i_4}
\e_{j_1 j_2 j_3 j_4} M^{i_1 j_1} M^{i_2 j_2} M^{i_3 j_3} N^{i_4} N^{j_4}
- P_i M^{ij} P_j \cr
& \qquad - R_\S R_\C - \Lambda_4^{12} \bigr] 
+ Y \bigr[ 2 N^i P_i + L_\S R_\C - L_\C R_\S \bigl]. }}
Checking global anomalies at the parton and hadron levels, we once again find 
that they all match in the $\Nf=4$ theory provided the Lagrange multipliers 
$X \sim (1; 1; 0, 0, 2)$ and $Y \sim (1; 1; -8, 0, 0)$ are included into the 
effective theory's spectrum.  

	The magnetic dual to the $SO(8)$ model has the symmetry group
\eqn\maggroupIII{\Gdual = SU(\Nf-3)_{\rm local} \times \bigl[ SU(\Nf) 
\times U(1)_\Y \times U(1)_\B \times U(1)_\R \bigr]_{\rm global}} 
%
%
and the rather complicated matter content
\eqn\magmatterIII{\eqalign{
q^\a_i &\sim \bigl( \fund; \antifund; 2,0,{5\Nf-18 \over \Nf(\Nf-3)} \bigr) \cr
\qp^\a &\sim \bigl( \fund; 1; 0,2, {\Nf-4 \over \Nf-3} \bigr) \cr
\qpp^\a &\sim \bigl( \fund; 1; 0,-2, {\Nf-4 \over \Nf-3} \bigr) \cr
\qbar_\a &\sim \bigl( \antifund; 1; -2\Nf, 0, -{\Nf-4 \over \Nf-3} \bigr) \cr
s_{\a\b} &\sim \bigl( \symbar; 1; 0,0,{2 \over \Nf-3}  \bigr) \cr
\ell_\S &\sim (1; 1; 2\Nf, 2,2 \bigr) \cr
\ell_\C &\sim (1; 1; 2\Nf, -2,2 \bigr) \cr
m^{(ij)} &\sim \bigl(1; \sym; -4, 0,2 - {12 \over \Nf} \bigr) \cr
n^i &\sim \bigl(1; \fund; 2\Nf-2, 0, 3 - {6 \over \Nf} \bigr).  \cr}}
Its tree level superpotential looks like
\eqn\WmagIII{
\Wmag = 2 n^i q^\a_i \qbar_\a + \ell_\S \qpp^\a \qbar_\a 
  - \ell_\C \qp^\a \qbar_\a 
  - m^{(ij)} q^\a_i s_{\a\b} q^\b_j 
  - \qp^\a s_{\a\b} \qpp^\b - \det s}
once we set to unity all dimensionful scales which multiply 
nonrenormalizable interaction terms.  It is straightforward to verify that 
the $SU(\Nf)^3$, $SU(\Nf)^2 U(1)_\Y$, $SU(\Nf)^2 U(1)_\B$, $SU(\Nf)^2 
U(1)_\R$, $U(1)_\Y$, $U(1)^3_\Y$, $U(1)_\B$, $U(1)^3_\B$, $U(1)_\R$, 
$U(1)^3_\R$, $U(1)^2_\Y U(1)_\B$, $U(1)^2_\Y U(1)_\R$, $U(1)^2_\B U(1)_\Y$, 
$U(1)^2_\B U(1)_\R$, $U(1)^2_\R U(1)_\Y$ and $U(1)^2_\R U(1)_\B$ global 
anomalies match between the electric and magnetic theories.  We also observe 
that the dual beta function coefficient
\eqn\magbetafuncIII{
\widetilde{b_0} = \cases{ 2, & $\Nf=5$ \cr
			  2 \Nf - 10, & $\Nf \ge 6$ \cr}} 
is positive everywhere throughout the magnetic theory's range of validity.  
The dual pair consequently exists in a nonabelian Coulomb phase for 
$5 \le \Nf < 16$.

	Operator maps between electric and magnetic gauge invariant operators 
are similar to those in the previous two $SO(7)$ models.  $L_\S$, $L_\C$, 
$M^{(ij)}$ and $N^i$ in \eleccompositesIII\ are mapped onto 
$\ell_\S$, $\ell_\C$, $m^{(ij)}$ and $n^i$ in \magmatterIII, while 
$P^{[ijk]}$, $R^{[ijkl]}_\S$ and $R^{[ijkl]}_\C$ are identified with the 
magnetic baryons 
\eqn\magbaryonsIII{\eqalign{
p^{[i_1 i_2 i_3]} &= \e^{i_1 \cdots i_\Nf} \e_{\a_1 \cdots \a_{\Nf-3}} 
  \sp q^{\a_1}_{i_4} \cdots q^{\a_{\Nf-3}}_{i_\Nf} \cr
r_\S^{[i_1 \cdots i_4]} &= \e^{i_1 \cdots i_\Nf} \e_{\a_1 \cdots \a_{\Nf-3}} 
  \sp q^{\a_1}_{i_5} \cdots q^{\a_{\Nf-4}}_{i_\Nf} \qp^{\a_{\Nf-3}} \cr
r_\C^{[i_1 \cdots i_4]} &= \e^{i_1 \cdots i_\Nf} \e_{\a_1 \cdots \a_{\Nf-3}} 
  \sp q^{\a_1}_{i_5} \cdots q^{\a_{\Nf-4}}_{i_\Nf} \qpp^{\a_{\Nf-3}}.  \cr}}
The electric baryons 
\eqn\elecbaryonsIII{\eqalign{
B_0^{[i_1 \cdots i_8]} &= \e^{\mu_1 \cdots \mu_8} V^{i_1}_{\mu_1} \cdots 
  V^{i_8}_{\mu_8} \cr
B_1^{a[i_1 \cdots i_6]} &= \e^{\mu_1 \cdots \mu_8} V^{i_1}_{\mu_1} \cdots
  V^{i_6}_{\mu_6} W^a_{\mu_7 \mu_8} \cr
B_2^{[i_1 \cdots i_4]} &= \e^{\mu_1 \cdots \mu_8} V^{i_1}_{\mu_1} \cdots 
V^{i_4}_{\mu_4} W^a_{\mu_5 \mu_6} W^a_{\mu_7 \mu_8} \cr}}
are matched onto the magnetic composites
\eqn\magantibaryonsIII{\eqalign{
b_0^{[i_1 \cdots i_8]} &= \e^{i_1 \cdots i_\Nf} \e^{\a_1 \cdots \a_{\Nf-3}} 
  \sp (sq)_{\a_1 i_9} \cdots (sq)_{\a_{\Nf-8} i_\Nf} (s \wtilde)_{\a_{\Nf-7}
\a_{\Nf-6}} (s \wtilde)_{\a_{\Nf-5} \a_{\Nf-4}} \qbar_{\a_{\Nf-3}} \cr
b_1^{a[i_1 \cdots i_6]} &= \e^{i_1 \cdots i_\Nf} \e^{\a_1 \cdots \a_{\Nf-3}} 
  \sp (sq)_{\a_1 i_7} \cdots (sq)_{\a_{\Nf-6} i_\Nf} (s \wtilde)_{\a_{\Nf-5}
\a_{\Nf-4}} \qbar_{\a_{\Nf-3}} \cr
b_2^{[i_1 \cdots i_4]} &= \e^{i_1 \cdots i_\Nf} \e^{\a_1 \cdots \a_{\Nf-3}} 
  \sp (sq)_{\a_1 i_5} \cdots (sq)_{\a_{\Nf-4} i_\Nf} \qbar_{\a_{\Nf-3}}. \cr}}
Finally, the exotic $SO(8)$ combinations of spinor, vector and gluon 
superfields
\eqn\elecexoticsIII{\eqalign{
C_0^{[i_1 \cdots i_7]} &= \e^{\mu_1 \cdots \mu_8} V^{i_1}_{\mu_1} \cdots 
  V^{i_7}_{\mu_7} Q_\S^\T \Gamma_{\mu_8} C Q_\C \cr
C_1^{a [i_1 \cdots i_5]} &= \e^{\mu_1 \cdots \mu_8} V^{i_1}_{\mu_1} \cdots 
  V^{i_5}_{\mu_5} W^a_{\mu_6 \mu_7} Q_\S^\T \Gamma_{\mu_8} C Q_\C \cr
C_2^{[i_1 i_2 i_3]} &= \e^{\mu_1 \cdots \mu_8} V^{i_1}_{\mu_1} V^{i_2}_{\mu_2} 
  V^{i_3}_{\mu_3} W^a_{\mu_4 \mu_5} W^a_{\mu_6 \mu_7} 
  Q_\S^\T \Gamma_{\mu_8} C Q_\C \cr}}
are associated with their $SU(\Nf-3)$ counterparts
\eqn\magexoticsIII{\eqalign{
c_0^{[i_1 \cdots i_7]} &= \e^{i_1 \cdots i_\Nf} \e^{\a_1 \cdots \a_{\Nf-3}} 
  \sp (sq)_{\a_1 i_8} \cdots (sq)_{\a_{\Nf-7} i_\Nf} (s \wtilde)_{\a_{\Nf-6}
\a_{\Nf-5}} (s \wtilde)_{\a_{\Nf-4} \a_{\Nf-3}} \cr
c_1^{a[i_1 \cdots i_5]} &= \e^{i_1 \cdots i_\Nf} \e^{\a_1 \cdots \a_{\Nf-3}} 
  \sp (sq)_{\a_1 i_6} \cdots (sq)_{\a_{\Nf-5} i_\Nf} (s \wtilde)_{\a_{\Nf-4}
\a_{\Nf-3}} \cr
c_2^{[i_1 i_2 i_3]} &= \e^{i_1 \cdots i_\Nf} \e^{\a_1 \cdots \a_{\Nf-3}} 
  \sp (sq)_{\a_1 i_4} \cdots (sq)_{\a_{\Nf-3} i_\Nf}. \cr }}
	
	Renormalization group flows induced by various deformations can be 
used to test the duality relationship between the $SO(8)$ and $SU(\Nf-3)$ 
models.  For example, we can integrate out all but four vector flavors.  The 
two quantum constraints encoded within the confining phase superpotential in 
\WconstraintIII\ are then recovered from the magnetic theory via semiclassical 
equations of motion in exactly the same fashion as we have seen in 
sections~2 and 3.  Here we will just consider adding a tree level mass term 
for the conjugate spinor field.  The electric theory then reduces to the 
Pouliot-Strassler $SO(8)$ model with $\Nf$ vectors and one spinor after 
the heavy field is integrated out \PouliotStrasslerI.  On the magnetic side, 
we supplement the superpotential in \WmagIII\ with $\Wtree = \mu \ell_\C$ and 
eliminate all fields involving the conjugate spinor.  From the equations 
of motion for $\ell_\C$ and $n^i$, we learn that the condensates 
\eqn\condensatesIII{\eqalign{
\vev{{q'}^\a} &= ( 0, \cdots, 0, \sqrt{\mu} ) \cr
\vev{\qbar_\a} &= (0, \cdots, 0, \sqrt{\mu} ) \cr
\vev{q^{\Nf-3}_i} &= 0 \cr}}
break the magnetic $SU(\Nf-3)$ gauge group down to $SU(\Nf-4)$.  
They also generate a bilinear mass term for ${q''}^{\Nf-3}$ and $\ell_\S$.
After some algebraic manipulations, we find 
\eqn\WmagIIIreduced{\Wmag \to - m^{(ij)} \qhat^\a_i \shat_{\a\b} \qhat^\b_j 
- \ell_\S \det \shat}
which is consistent with the magnetic superpotential expression in 
ref.~\PouliotStrasslerI.

\newsec{$\pmb{SO(9)}$ with $\pmb{\Nf}$ vectors and one spinor}

	The final orthogonal group supersymmetric gauge theory which we 
investigate is based upon 
\eqn\symgroupIV{G = SO(9)_{\rm local} \times \bigl[ SU(\Nf) \times U(1)_\Y 
\times U(1)_\R \bigr]_{\rm global}}
and contains the matter fields 
\eqn\matterIV{\eqalign{
V^i_\mu & \sim \bigl( 9; \fund ; -2,1-{5 \over \Nf} \bigr) \cr
Q^\A & \sim \bigl( 16; 1 ; \Nf, 0 \bigr). \cr}}
As both the 9-dimensional vector and 16-dimensional spinor irreps 
of $SO(9)$ are real, this model is nonchiral.  It is also asymptotically 
free provided it contains $\Nf < 19$ vector flavors so that its 
Wilsonian beta function coefficient $b_0 = 19 - \Nf$ is positive and its full 
beta function is negative.

	At generic points in moduli space, the model follows the gauge 
symmetry breaking pattern \Elashvili
\eqn\patternIV{SO(9) \> {\buildrel 16 \over \longrightarrow} \>
	       SO(7) \> {\buildrel 9 \over \longrightarrow} \>
               G_2 \> {\buildrel 9 \over \longrightarrow} \> 
               SU(3) \> {\buildrel 9 \over \longrightarrow} \>
               SU(2) \> {\buildrel 9 \over \longrightarrow} \>
               1.}
The first link in this chain indicates that the color group breaks to 
$SO(7)$ along the spinor flat direction.  As we shall see, the electric 
$SO(9)$ theory and its magnetic counterpart reduce to Pouliot's original 
chiral-nonchiral dual pair along this direction \PouliotStrasslerII.  Taking 
into account the symmetry breaking information in \patternIV\ as well as the 
spinor tensor product decomposition
\eqn\soninespinorprod{16 \times 16 = [0]_\S + [1]_\S + [2]_\A + [3]_\A + 
[4]_\S,}
we form the composite operators
\eqn\eleccompositesIV{\eqalign{
L &= Q^\T C Q 
	\sim \bigl(1; 1; 2\Nf , 0 \bigr) \cr
M^{(ij)} &= (V^\T)^{i \mu} V^j_\mu 
	\sim \bigl(1; \sym; -4, 2-{10 \over \Nf} \bigr) \cr
N^i &= Q^\T \Vslash^i C Q 
	\sim \bigl(1; \fund; 2 \Nf - 2, 1 - {5\over \Nf} \bigr) \cr
R^{[ijkl]} &= {1 \over 4!} Q^\T \Vslash^{[i} \Vslash^j \Vslash^k \Vslash^{l]}
C Q 
	\sim \bigl(1; \antifour; 2\Nf - 8; 4 - {20\over\Nf} \bigr) \cr 
T^{[ijklm]} &= {1 \over 5!} Q^\T \Vslash^{[i} \Vslash^j \Vslash^k \Vslash^l
\Vslash^{m]} C Q 
	\sim \bigl(1; \antifive; 2\Nf - 10; 5 - {25 \over \Nf} \bigr). \cr }}
Simple degree of freedom counting demonstrates that these hadrons represent 
all massless modes in the $\Nf \le 4$ models.  On the other hand, the hadron 
count exceeds the number of independent scalar potential flat directions by 
two when $\Nf=5$.  The operators in \eleccompositesIV\ are then 
constrained by the two quantum relations contained within
\eqn\WconstraintIV{\eqalign{
W_{\Nf=5} &= X \bigl[ L^2 \det M - {1 \over 4!} \e_{i_1 i_2 i_3 i_4 i_5}
\e_{j_1 j_2 j_3 j_4 j_5} M^{i_1 j_1} M^{i_2 j_2} M^{i_3 j_3} M^{i_4 j_4}
N^{i_5} N^{j_5} \cr
& \qquad - R_i M^{ij} R_j + T^2 - \Lambda_5^{14} \bigr] + 
Y \bigl[ N^i R_i - L T \bigr]. \cr}}

	In order to check the validity of these two $\Nf=5$ quantum 
constraints, it is instructive to determine how they decompose 
along the spinor flat direction.  To begin, we need to explicitly embed the 
$SO(7)$ subgroup inside $SO(9)$.  Working with the $16 \times 16$ Gamma 
matrices 
\eqn\Gammamatrices{\eqalign{
\Gamma_1 &= \sigtwo \times \sigthree \times \sigthree \times \sigthree\cr
\Gamma_2 &= -\sigone \times \sigthree \times \sigthree \times \sigthree \cr
\Gamma_3 &= 1 \times \sigtwo \times \sigthree \times \sigthree \cr}
\qquad
\eqalign{
\Gamma_4 &= -1 \times \sigone \times \sigthree \times \sigthree \cr
\Gamma_5 &= 1 \times 1 \times \sigtwo \times \sigthree \cr
\Gamma_6 &= -1 \times 1 \times \sigone \times \sigthree \cr}
\qquad
\eqalign{
\Gamma_7 &= 1 \times  1 \times 1 \times \sigtwo \cr
\Gamma_8 &= -1 \times  1 \times 1 \times \sigone \cr
\Gamma_{9} &= \sigthree \times \sigthree \times \sigthree \times\sigthree,\cr}}
we first form the spinor irrep generators $ M_{\mu\nu} = - {i \over 4} \bigl[ 
\Gamma_\mu, \Gamma_\nu \bigr]$ and then take $M_{12} \equiv H_1$, $M_{34} 
\equiv H_2$, $M_{56} \equiv H_3$ and $M_{78} \equiv H_4$ as the four members 
of the $SO(9)$ Cartan subalgebra.  We next act upon $Q$ with 
\eqn\sosevencartangens{\eqalign{
h_1 &= H_1 + H_2 - H_3 - H_4 \cr
h_2 &= H_1 - H_2 - H_3 + H_4 \cr
h_3 &= H_1 - H_2 + H_3 - H_4 \cr}}
and find that the flat direction vev $\vev{Q} = ( a, 0, \cdots, 0, a)$
is left invariant.  We also recover the correct 
$SO(9) \> {\buildrel \vev{16} \over \longrightarrow} \> SO(7)$ 
branching rule $16 \to 8 + 7 + 1$ for the spinor field.  The 
linear combinations in \sosevencartangens\ are thus identified as the three 
generators of the $SO(7)$ Cartan subalgebra.  

	After computing the eigenvalues and eigenvectors of $h_1$, $h_2$ and 
$h_3$ in the vector irrep, we deduce that the $9$ of $SO(9)$ decomposes into 
$8+1$ under $SO(7)$ as
\eqn\ninedecomp{V_\mu = \pmatrix{ 
q_1 - i q_8 \cr
i q_1 - q_8 \cr
- q_4 - i q_5 \cr
- i q_4 - q_5 \cr
q_7 + i q_2 \cr
i q_7 + q_2 \cr
q_6 - i q_3 \cr
i q_6 - q_3 \cr
\phi \cr}.}
Symmetry considerations fix only the relative phases for the spinor components 
appearing inside this vector.  We choose their absolute phases so that the 
$SO(9)$ hadrons break apart simply into the $SO(7)$ mesons 
$m^{(ij)} = {q^i}^\T c \, q^j$ and baryons 
$b_i = {1 \over 4!} \e_{ijklm} {q^j}^\T \gamma_\mu c \, q^k 
{q^\ell}^\T \gamma^\mu c \, q^m$: 
\eqn\hadrondecomp{\eqalign{
L & \to -2 a^2 \cr
M^{(ij)} & \to 2 i m^{(ij)} + \phi^i \phi^j \cr
N^i & \to - 2 a^2 \phi^i \cr
R_i & \to 8 a^2 b_i \cr
T & \to 8 a^2 b_i \phi^i. \cr}}
Inserting these field decompositions into \WconstraintIV, we recover the 
quantum relation $\det m - b_i m^{ij} b_j = \Lambda_7^{10}$ in Pouliot's model 
from the first constraint in $W_{\Nf=5}$ \Pouliot, while the second 
identically vanishes.  These results provide strong consistency checks 
on the $\Nf=5$ $SO(9)$ theory.  

	The ground state structure of the $\Nf \gtap 6$ $SO(9)$ model can 
be simply understood in terms of its dual based upon 
\eqn\maggroupIV{\Gdual = SU(\Nf-4)_{\rm local} \times \bigl[ SU(\Nf) 
\times U(1)_\Y \times U(1)_\R \bigr]_{\rm global}. }
The magnetic theory has chiral matter content
\eqn\magmatterIV{\eqalign{
q^\a_i &\sim \bigl( \fund; \antifund; 2, {4 \over \Nf} 
  {\Nf-5 \over \Nf-4} \bigr) \cr
\qp^\a &\sim \bigl( \fund; 1; 0, {\Nf-5 \over \Nf-4} \bigr) \cr
\qbar_\a &\sim \bigl( \antifund; 1; -2\Nf, {\Nf-3 \over \Nf-4} \bigr) \cr
s_{\a\b} &\sim \bigl( \symbar; 1; 0, {2 \over \Nf-4} \bigr) \cr
\ell &\sim (1; 1; 2\Nf, 0 \bigr) \cr
m^{(ij)} &\sim \bigl(1; \sym; -4, 2 - {10 \over \Nf} \bigr) \cr
n^i &\sim \bigl(1; \fund; 2\Nf-2, 1 - {5 \over \Nf} \bigr) \cr}}
and tree level superpotential
\eqn\WmagIV{
\Wmag = n^i q^\a_i \qbar_\a - \ell \qp^\a \qbar_\a 
  - m^{(ij)} q^\a_i s_{\a\b} q^\b_j + \qp^\a s_{\a\b} \qp^\b - \det s.}
Its magnetic beta function 
\eqn\magbetafuncIV{
\widetilde{b_0} = \cases{ 4, & $\Nf=6$ \cr
			  2 \Nf - 12, & $\Nf \ge 7$ \cr}} 
is positive for $6 \le \Nf < 19$ flavors.  The theory then exists at the 
origin of moduli space in a nonabelian Coulomb phase.  All the usual comments 
regarding anomaly matching, global symmetry agreement and scale simplification 
hold for this dual like its predecessors.  The maps between composite 
operators in the electric and magnetic theories also follow the familiar 
pattern.  The $R^{[ijkl]}$ and $T^{[ijklm]}$ hadrons are identified with 
\eqn\magbaryonsIV{\eqalign{
r^{[i_1 \cdots i_4]} &= \e^{i_1 \cdots i_\Nf} \e_{\a_1 \cdots \a_{\Nf-4}} 
 \sp q^{\a_1}_{i_5} \cdots q^{\a_{\Nf-4}}_{i_\Nf} \cr
t^{[i_1 \cdots i_5]} &= \e^{i_1 \cdots i_\Nf} \e_{\a_1 \cdots \a_{\Nf-4}} 
  \sp q^{\a_1}_{i_6} \cdots q^{\a_{\Nf-5}}_{i_\Nf} \qp^{\a_{\Nf-4}}, \cr}}
while the electric baryons and exotics 
\eqn\elecbaryonsIV{\eqalign{
B_0^{[i_1 \cdots i_9]} &= \e^{\mu_1 \cdots \mu_9} V^{i_1}_{\mu_1} \cdots 
  V^{i_9}_{\mu_9} \cr
B_1^{a[i_1 \cdots i_7]} &= \e^{\mu_1 \cdots \mu_9} V^{i_1}_{\mu_1} \cdots
  V^{i_7}_{\mu_7} W^a_{\mu_8 \mu_9} \cr
B_2^{[i_1 \cdots i_5]} &= \e^{\mu_1 \cdots \mu_9} V^{i_1}_{\mu_1} \cdots 
V^{i_5}_{\mu_5} W^a_{\mu_6 \mu_7} W^a_{\mu_8 \mu_9} \cr
C_0^{[i_1 \cdots i_8]} &= \e^{\mu_1 \cdots \mu_9} V^{i_1}_{\mu_1} \cdots 
  V^{i_8}_{\mu_8} Q^\T \Gamma_{\mu_9} C Q \cr
C_1^{a [i_1 \cdots i_6]} &= \e^{\mu_1 \cdots \mu_9} V^{i_1}_{\mu_1} \cdots 
  V^{i_6}_{\mu_6} W^a_{\mu_7 \mu_8} Q^\T \Gamma_{\mu_9} C Q \cr
C_2^{[i_1 \cdots i_4]} &= \e^{\mu_1 \cdots \mu_9} V^{i_1}_{\mu_1} \cdots 
  V^{i_4}_{\mu_4} W^a_{\mu_5 \mu_6} W^a_{\mu_7 \mu_8} 
  Q^\T \Gamma_{\mu_9} C Q \cr}}
are matched onto the magnetic operators
\eqn\magexoticsIV{\eqalign{
b_0^{[i_1 \cdots i_9]} &= \e^{i_1 \cdots i_\Nf} \e^{\a_1 \cdots \a_{\Nf-4}} 
  \sp (sq)_{\a_1 i_{10}} \cdots (sq)_{\a_{\Nf-9} i_\Nf} (s \wtilde)_{\a_{\Nf-8}
\a_{\Nf-7}} (s \wtilde)_{\a_{\Nf-6} \a_{\Nf-5}} \qbar_{\a_{\Nf-4}} \cr
b_1^{a[i_1 \cdots i_7]} &= \e^{i_1 \cdots i_\Nf} \e^{\a_1 \cdots \a_{\Nf-4}} 
  \sp (sq)_{\a_1 i_8} \cdots (sq)_{\a_{\Nf-7} i_\Nf} (s \wtilde)_{\a_{\Nf-6}
\a_{\Nf-5}} \qbar_{\a_{\Nf-4}} \cr
b_2^{[i_1 \cdots i_5]} &= \e^{i_1 \cdots i_\Nf} \e^{\a_1 \cdots \a_{\Nf-4}} 
  \sp (sq)_{\a_1 i_6} \cdots (sq)_{\a_{\Nf-5} i_\Nf} \qbar_{\a_{\Nf-4}} \cr 
c_0^{[i_1 \cdots i_8]} &= \e^{i_1 \cdots i_\Nf} \e^{\a_1 \cdots \a_{\Nf-4}} 
  \sp (sq)_{\a_1 i_9} \cdots (sq)_{\a_{\Nf-8} i_\Nf} (s \wtilde)_{\a_{\Nf-7}
\a_{\Nf-6}} (s \wtilde)_{\a_{\Nf-5} \a_{\Nf-4}} \cr
c_1^{a[i_1 \cdots i_6]} &= \e^{i_1 \cdots i_\Nf} \e^{\a_1 \cdots \a_{\Nf-4}} 
  \sp (sq)_{\a_1 i_7} \cdots (sq)_{\a_{\Nf-6} i_\Nf} (s \wtilde)_{\a_{\Nf-5}
\a_{\Nf-4}} \cr
c_2^{[i_1 \cdots i_4]} &= \e^{i_1 \cdots i_\Nf} \e^{\a_1 \cdots \a_{\Nf-4}} 
  \sp (sq)_{\a_1 i_5} \cdots (sq)_{\a_{\Nf-4} i_\Nf}. \cr }}

	Just as it is possible to recover the confining phase quantum 
constraint in Pouliot's $SO(7)$ model starting from the $SO(9)$ theory, so 
can its nonabelian Coulomb phase be reproduced as well.  When the spinor vev 
breaks $SO(9)$ down to $SO(7)$, the $\ell$, $m^{(ij)}$ and $n^i$ fields 
appearing in the first three terms of the magnetic superpotential need to be 
replaced by their decomposed forms listed in \hadrondecomp:
\eqn\WmagIVa{
\Wmag \to -2 a^2 \phi^i q^\a_i \qbar_\a + 2 a^2 \qp^\a \qbar_\a 
  - \bigl[ 2 i m^{(ij)} + \phi^i \phi^j \bigr] q^\a_i s_{\a\b} q^\b_j 
  + \qp^\a s_{\a\b} \qp^\b - \det s.}
Recalling that Pouliot's model contains only spinor matter, we render all the 
$\phi^i$ singlets heavy by introducing additional singlets $\chi_i$ and 
supplementing $\Wmag$ with the mass term $\Wtree = \mu \phi^i \chi_i$. 
The $\chi_i$ equation of motion freezes $\phi^i$ at zero expectation 
value.  We must also integrate out $\qp^\a$ and $\qbar_\a$ since both acquire 
mass from the bilinear term in \WmagIVa.  The magnetic theory's particle 
content and superpotential 
\eqn\WPouliot{\Wmag \to - 2 i m^{(ij)} q^\a_i s_{\a\b} q^\b_j - \det s}
thus reduce to those of Pouliot's SO(7) dual along the spinor flat direction 
\Pouliot.

\newsec{Conclusion}

	The chiral-nonchiral dual pairs which we have examined in this 
article are related to one another by renormalization group flows along 
various flat directions.  We display the web which connects these four theories 
to one another as well as to other dual pairs in fig.~1.  As can be seen in the
figure, Higgsing the electric theory induces mass decoupling in its 
magnetic counterpart and vice-versa.  This basic duality feature was first 
noted by Seiberg in his dual to supersymmetric QCD \SeibergI.  We have 
repeatedly observed that the complete breaking of the magnetic gauge group 
and the resulting condensation of magnetic monopoles are correlated with the 
onset of confinement in the electric theory.   This phenomenon is also 
familiar from SUSY QCD \SeibergI.  However, our particular $SO(7)$, $SO(8)$ 
and $SO(9)$ models possess no analogue of the quantum moduli 
space that exists in $\Nf=\Nc+1$ SUSY QCD.  Instead, their nonabelian Coulomb 
phases give way to confinement characterized by nontrivial quantum relations 
among colorless moduli.  As we have seen, these electric theory 
quantum constraints can be recovered from semiclassical equations of motion 
in the magnetic theories.  So duality relates complicated nonperturbative 
phenomena in one description to weakly coupled results in the other.

	Several extensions of this work would be interesting to pursue.  
One could continue to explore further dual pairs which arise along various 
flat directions of the Pouliot-Strassler $SO(10)$ model.  Distler and Karch 
have recently investigated the long distance emergence of accidental global 
symmetries in duals to $SO(5)$, $SO(3)$ and $SU(3)$ electric theories that
appear further down on the web in fig.~1 \Distler.  Studies similar to theirs 
and ours could be undertaken for other examples of chiral-nonchiral dual 
pairs.  But clearly, it would be preferable to generalize all these results 
and find new duals which reduce to those displayed in the figure as special 
cases.  Obvious starting points to consider include $SO(9)$ and $SO(10)$ with 
two spinors and $\Nf$ vector flavors as well as $SO(11)$ with one spinor and 
various numbers of vectors \Cho.  As we noted in our opening remarks, there 
unfortunately exists no systematic field theory method for deducing magnetic 
counterparts to these larger electric theories.  But they must at least 
reduce to those which we have analyzed in this paper along certain flat 
directions.  It seems likely that such bigger dual pairs also extend the 
patterns and trends which we repeatedly encountered in this work.  So we hope 
our findings will provide useful clues in future duality searches.

\bigskip\bigskip\bigskip\bigskip
\centerline{{\bf Acknowledgments}}
\bigskip

        It is a pleasure to thank Howard Georgi, Matt Strassler and 
especially Per Kraus for many helpful discussions. 

\listrefs
\bye